\documentclass[journal,a4paper]{IEEEtran}
\usepackage[compress]{cite}
\usepackage{color,soul}
\usepackage{tabularx,ragged2e,booktabs}
\ifCLASSINFOpdf
   \usepackage[pdftex]{graphicx}
    \usepackage{epstopdf}
  \graphicspath{{figures/},{Figures/}}
  \DeclareGraphicsExtensions{.pdf,.eps,.png}
\else
   \usepackage[dvips]{graphicx}
  \graphicspath{{figures/},{Figures/}}
  \DeclareGraphicsExtensions{.eps,.pdf}
\fi
\usepackage[cmex10]{amsmath}
\usepackage{nicefrac}
\usepackage{textcomp}
\usepackage{siunitx}
\usepackage{comment}
\usepackage[compress,capitalize]{cleveref}
\crefname{figure}{Fig.}{Figs.}
\crefname{equation}{}{}
\crefname{secinapp}{Appendix}{Appendices}
\usepackage{microtype}
\include{OperatorsAndAbbreviations}
\begin{document}
\title{Charge-based Model for Ultra-Thin Junctionless DG FETs, Including Quantum Confinement }
\author{Majid Shalchian, Farzan Jazaeri, and Jean-Michel~Sallese
 
\thanks{Manuscript received March 26, 2018.}%
            
\thanks{
Majid Shalchian is with the Electrical Engineering Department, Amirkabir University of Technology,  424 Hafez Ave, Tehran, Iran, 15875-4413, (e-mail: shalchian@aut.ac.ir). Farzan Jazaeri and Jean-Michel Sallese are respectively with the Integrated Circuits Laboratory (ICLAB) and Electron Device Modeling and Technology Laboratory (EDLAB) of the Ecole Polytechnique F\'{e}d\'{e}rale de Lausanne (EPFL), Neuch\^{a}tel, Switzerland.}%
}%
\markboth{IEEE TRANSACTIONS ON ELECTRON DEVICES,~Vol.~XX, No.~XX, X~XXXX}%
{Shell \MakeLowercase{\textit{et al.}}: Bare Demo of IEEEtran.cls for IEEE Journals}
 
\maketitle
 
\begin{abstract}
This paper presents a generalization of the charge-based model for ultra-thin junctionless double-gate FETs \cite{5872019, Book} by including quantum electron density. The analytical derivation relies on a first order correction to the infinite quantum well. When restricting the analysis to the first and second quantized states, the free carrier charge distribution and the current in an ultra-thin body junctionless double gate FETs is in agreement with numerical TCAD simulations in all the regions of operation, i.e. from deep depletion to accumulation and from linear to saturation.
 
\end{abstract}
 
\begin{IEEEkeywords}
        Junctionless FETs, double-gate FETs, nanowire FETs, quantum well, UTBSOI.
\end{IEEEkeywords}

\section{Introduction}
\IEEEPARstart{J}{unctionless} double-gate and nanowires field-effect transistors are among viable candidates for the next generation of digital and analog applications \cite{Nature, 079411, 5934399}.
In junctionless double-gate FET (JLDG), the channel is uniformly doped from the source to the drain and is controlled by two gates \cite{Nature, 3079411}.
These devices have no source-drain pn junction, a feature that  relaxes some critical processing steps, an advantage over the conventional MOSFETs \cite{JAZAERI2013103}.
 
A generic charge-based model was proposed for symmetric double-gate structures \cite{5872019,6879262, Book}. In that derivation, non-degenerate Boltzmann statistic and 3D density of states were used to obtain the surface electric-field and the surface potential in the channel. The model predicts accurately DC and AC electrical behaviors from depletion to accumulation \cite{5872019,Book, 6642051, 6525384}, but neglects channel quantization, which is not a valid assumption for silicon layers below \SI{10}{nm}. Similarly, at high gate overdrive voltages, the Boltzmann statistics combined with the 2D density of states tends to overestimate the free carrier densities and should be replaced with the more general Fermi-Dirac statistics. A quantum model was proposed in \cite{6151818}, but the analysis was restricted to deep depletion where the mobile charge density is low and the silicon channel is almost fully depleted. Based on this assumption, the contribution of the mobile charge density in the Poisson equation was neglected, meaning that the model is inaccurate above the threshold. Another approach \cite{1674-4926-36-2-024001,Chanda2015} is to incorporate the effect of charge quantization as a shift in the I-V characteristics, however due to the coupling between Poisson and Schr\"{o}dinger equations, the amount of the shift varies with the gate voltage.   
\begin{figure}[!t]
                \vspace{0.2cm}
                \centering
                \includegraphics[width=0.9\columnwidth]{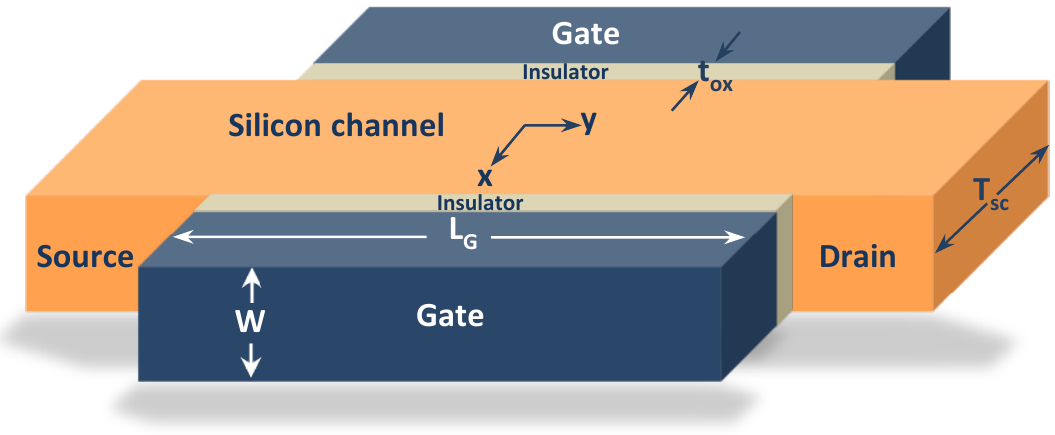}
                \caption{Schematic view of an \textit{n}-type double-gate junction FET investigated.}
                \label{Fig1}
\end{figure}
The aim of this paper is to generalize the charge-based model in \cite{5872019, Book} to include quantum confinement and Fermi-Dirac Statistics in ultra-thin junctionless double-gate FETs. The paper is organized as follows: Section II defines the structure and core equations, section III derives approximate solution for the charge densities, section IV addresses calculation of the current versus gate voltage. The results and validity of the assumptions are discussed in section V. Finally, conclusions are drawn in Sec. VI.
\section{Device Structure and Model Derivation} 
\begin{figure}[!t]
                \centering
                \includegraphics[width=0.65\columnwidth]{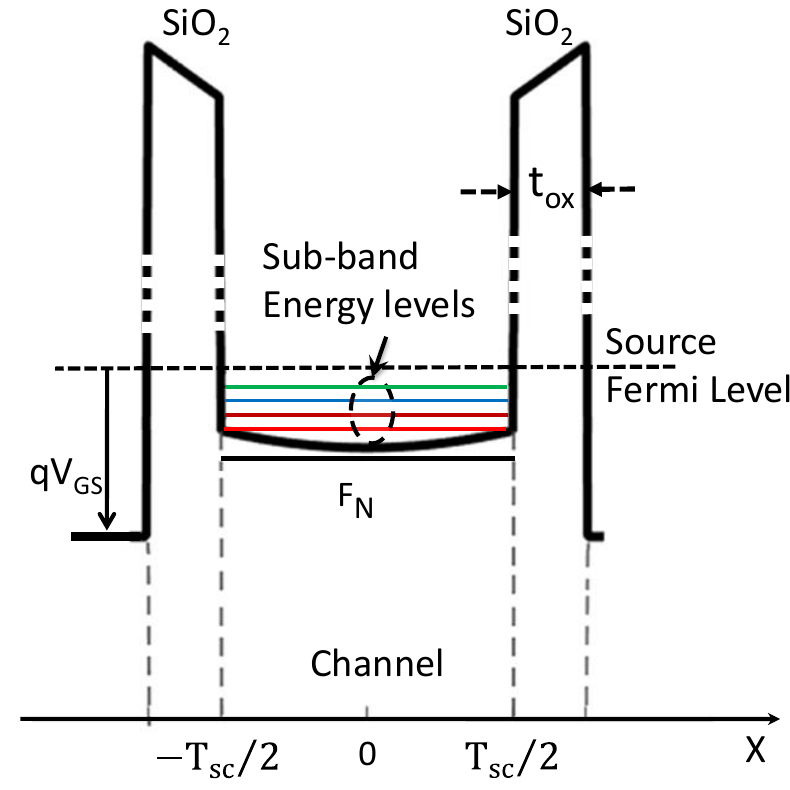}
                \caption{Sketch of the energy diagram for the junctionless doped DG MOSFET.}
                \label{Fig2}
\end{figure}

\begin{figure}[!b]
                \centering
                \includegraphics[width=1\columnwidth]{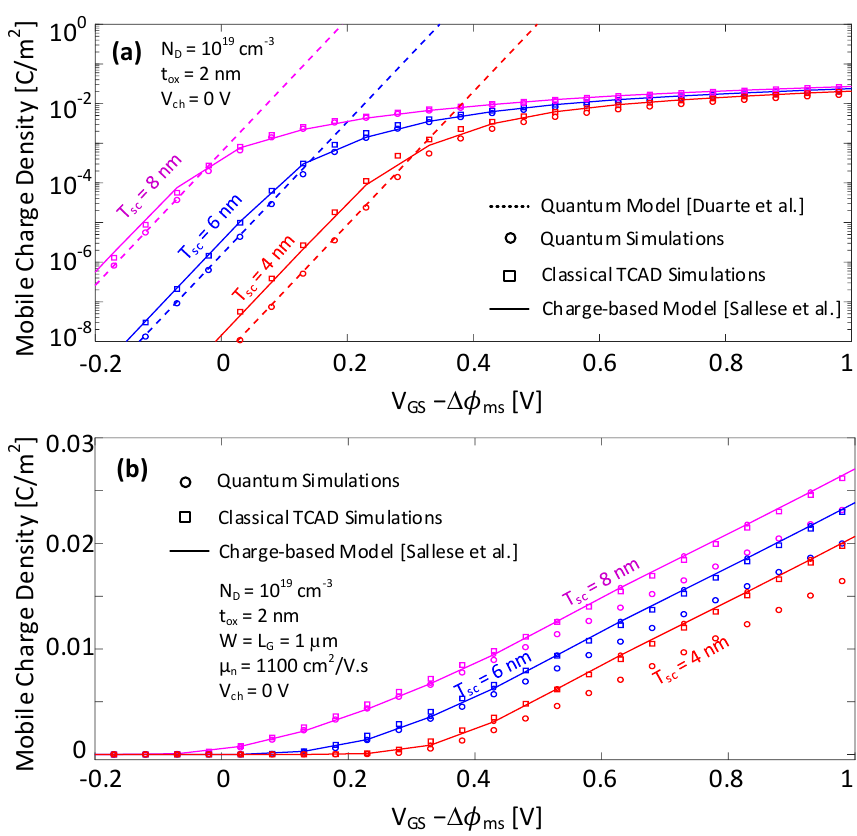}
                \caption{Mobile charge density versus the effective gate voltage for different $ T_{sc} $ in logarithmic (a) and linear scales (b). Classic charge-based model \cite{5872019, Book}: dashed lines with circle symbol, quantum model in deep depletion mode: dotted lines, TCAD simulation based on classic Boltzmann statistics: square symbols, and quantum TCAD simulation based on Poisson- Schr\"{o}dinger: triangle symbols. $ N_D$ = \SI{E19}{cm^{-3}}, $ t_{ox}$ = \SI{2}{nm}, $ W $ = \SI{1}{\micro m}, $ L_G$ = \SI{1}{\micro m}, $ \mu_n$ = \SI{1100}{cm^2/V\,s}, and $ V_{ch}$ = \SI{0}{V}.          
                }
                \label{Fig3}
\end{figure}

The structure of a n-type junctionless double-gate FET and the energy band diagram under a positive gate-to-source voltage is depicted in Fig. \ref{Fig1} and Fig. \ref{Fig2} respectively. The device parameters are listed in table \ref{Table}. The mode of operation of the JLFET can be either depletion, accumulation, or hybrid \cite{5872019, Book}. When the whole channel operates in depletion, the gate-to-source voltage satisfies $ V_{GS} < V_{FB,S} $, where $ V_{FB,S} $ is the flat-band condition at the source given by $ V_{FB,S}= \Delta\phi_{ms}+U_T\ln(N_D/n_i) $. When the whole channel is in accumulation, $ V_{GS} > V_{FB,D} $, where $ V_{FB,D} $ is the flat-band condition at the drain given by $ V_{FB,D}= V_{DS}+\Delta\phi_{ms}+U_T\ln(N_D/n_i) $. In between is the hybrid state where depletion and accumulation coexist (respectively towards the drain and the source).
In Fig. \ref{Fig3}, the mobile charge density obtained from the classical charge-based model in \cite{5872019, Book} agrees with TCAD simulations when quantum corrections are neglected. However, the same model shows some mismatch when quantization is included. This is evidenced in ultra-thin channel layers (i.e. \SI{4}{nm} \SI{6}{nm}, and \SI{8}{nm}) when  TCAD simulations include quantum effect. Such quantum corrections were first introduced in \cite{6151818}, but still, the derivation was only valid in the subthreshold region.
Moreover, Baccarani et. al \cite{777154} demonstrated that for ultra-thin channels, discrete sub-band energy states can be solved using a first order correction based on the time-independent perturbation theory of the Schr\"{o}dinger equation. The discrete sub-band energy states were given as follows;
\begin{equation} \label{Enk} E_{n,k}=E_{co}+\dfrac{(n\pi\hbar)^2}{2m^*_{c,k}T^2_{sc}}+ \dfrac{qQ_{sc}T_{sc}}{24 \varepsilon_{si}}\left[1-\dfrac{6}{(n\pi)^2}\right],
\end{equation}\\
where $ n $ and $ k $ are respectively the sub-band number and  valley number. For silicon with $<100>$ orientation, two valleys are considered. The lower valley $ (k=1) $ twofold degenerated ($ g_1=2 $) with $ m^*_{c,1}=0.92m_0 $, and the higher valley $(k=2)$ fourfold degenerated ($ g_2=4 $) with $ m^*_{c,2}=0.19m_0 $ ($ m_0 $ is the free electron mass). The total semiconductor charge density per unit area $ Q_{sc} $ is given by
\begin{equation}
\label{Q_{sc}} Q_{sc}=Q^{QM}_{2D}+Q_{fix}, 
\end{equation}
where $Q^{QM}_{2D}$ and  $Q_{fix}$ $(Q_{fix}= qN_{D}T_{sc}) $ are respectively the (quantized) mobile charge density and the fixed charge density per unit area. {\color{black} Despite relation (\ref{Enk}) was  obtained for lightly doped channels \cite{777154}, its validity was demonstrated for highly doped channel JL devices\cite{6151818} in the subthreshold region. In our model, we propose to use (\ref{Enk}) also above the threshold.
 
Fig. \ref {Fig4} (a) shows that the lowest subband energy level as a function of $T_{sc}$ in the subthreshold region predicted by (\ref{Enk}) closely follows numerical simulations.  As illustrated in Fig. \ref{Fig4} (b), two regions of operation are evidenced. When the effective gate voltage is below $0.3$ V, the mobile charge density is negligible (see Fig \ref {Fig3} (b)), and the curvature of the potential well is dominated by the ionized donors, i.e. subband energies are gate voltage independent. When the effective gate voltage is above $0.3$ V, the negative mobile charge density becomes comparable to the positive fixed charge density and $Q_{sc}$ decrease with the gate voltage. This means that the curvature in the potential decreases, as for the correction of the ideal infinite quantum well. According to the model, the third term in RHS of (\ref{Enk}) is proportional to $Q_{sc}$, which is responsible for a decrease in subbands energy with an increase in the gate voltage. This is comparable to the reduction of subband energies upon widening of the effective potential well as reported in \cite{doi:10.1063/1.2907330} for accumulation mode transistors. Fig. \ref {Fig4}(b) also indicates that the first order correction proposed in \cite{777154} remains valid even above the threshold, including accumulation.
The quantized mobile charge density per unit area, $ Q^{QM}_{2D} $ is the sum of charges in quantized sub-bands given by \cite{777154}
\begin{table}[!t]
                \centering
                \renewcommand{\arraystretch}{1.2}
                \caption[Physical parameters of a junctionless double-gate FET.]{Physical parameters of junctionless double-gate FET.}
                \label{Table}
                \scalebox{1.1}{
                                \begin{tabular}{lll}
                                                \hline
                                                \textbf{Parameter} & \textbf{Symbol} & \textbf{Value} \\
                                                \hline
                                                \textbf{Channel Thickness} & $ T_{sc} $ & \SI{4}{nm}, \SI{6}{nm}, \SI{8}{nm} \\
                                                \textbf{Channel Doping} & $ N_D $ & \SI {E+19}{cm^{-3}}  \\
                                                \textbf{Oxide Thickness} & $ t_{ox} $ & \SI{2}{nm} \\
                                                \textbf{Channel Length} & $ W $ & \SI{1}{$ \mu $m} \\
                                                \textbf{Channel Width} & $ L_G $ & \SI{1}{$ \mu $m} \\
                                                \textbf{Work-function Difference}  & $ \Delta \phi_{ms} $ & \SI{0}{V} \\
                                                \textbf{Electron Mobility}  & $ \mu_n $ & \SI{1100}{cm^{2}/V.s } \\
                                                \textbf{Silicon Permittivity}  & $ \varepsilon_{si} $ & 11.68$ \varepsilon_{0} $  \\
                                                \textbf{Silicon Dioxide Permittivity}  & $ \varepsilon_{ox} $ & 3.9$ \varepsilon_{0} $  \\
                                                \hline
                \end{tabular}}
\end{table}
\begin{equation} \label{Q}
\begin{split}
&Q^{QM}_{2D}=\sum_{n=1}^{N_t}Q_{bn}\\
&=-q\sum_{n=1}^{N_t}\sum_{k=1}^{2} g_kN_{k}\ln\left[1+\exp\left(\dfrac{-E_{n,k}+E_{Fn}}{k_BT}\right)\right]\\
&=\!-q\!\sum_{n=1}^{N_t}\!\sum_{k=1}^{2}g_kN_{k} \ln\!\left[1\!+\!\exp\!\left(\!\dfrac{-E_{n,k}+\!E_{c0}-\!E_{c0}+\!E_{Fn}}{k_BT}\!\right)\right]
\end{split}
\end{equation}
{\color{black} where $ N_k= m^*_{d,k}k_B T/\pi\hbar^2$ is the 2D effective density of states in the subband } at energy $ E_{n,k} $ and $ N_t $ is the number of subbands. $ m_{d,1}^*=0.19 m_0 $ and $ m_{d,2}^*=0.417 m_0 $ are density of states effective masses for valleys $ 1 $ and $2 $ respectively \cite{Taur}. $ E_{Fn} $ is the electron quasi Fermi potential related to channel potential $(V_{ch})$ by
\begin{equation} \label{E_{fn}}
E_{fn}-E_{c0}=q\psi_0-qV_{ch}+k_{B}T\ln\left(\dfrac{n_i}{N_c}\right),
\end{equation}
\begin{table}[!b]
                \centering
                \renewcommand{\arraystretch}{2.2}
                \caption[Percentage of Electric Charge Density in Different Subbands at the Flatband Condition.]{Percentage of Electric Charge Density in Different Subbands at the Flatband Condition.}
                \label{Table2}
                \scalebox{0.73}{
                                \begin{tabular}{lcccc}
                                                \hline
                                                \normalsize{\textbf{Channel Thickness}} & \normalsize{\textbf{Subband1}} & \normalsize{\textbf{Subband-2}}& \normalsize{\textbf{Subband-3}} & \normalsize{\textbf{Other Subbands}}\\
                                                \hline
                                                \normalsize{\textbf{$ T_{sc}= 8 nm$}} & \normalsize{78.4}\% & \normalsize{15.9}\%&  \normalsize{4.6}\%& \normalsize{1.1}\%\\
                                                \normalsize{\textbf{$ T_{sc}= 6 nm $}} & \normalsize{84.6}\% & \normalsize{13.6}\%&  \normalsize{1.6}\%& \normalsize{0.2}\%\\
                                                \normalsize{\textbf{$ T_{sc}= 4 nm$}} & \normalsize{94.8}\% & \normalsize{5.15}\%&  \normalsize{0.04}\%& \normalsize{0.01}\%\\
                                                \hline
                \end{tabular}}
\end{table}
where $ E_{c0} $ is conduction band edge at $ x=0 $, $ \psi_0 $ is the central potential of the channel, $n_i$ is intrinsic carrier density and  $N_c$ is the conduction band effective density of states. Under the assumption that only the first and second lowest sub-bands ($ n=1,2 $) concentrate most of the mobile charge density in practical situations, the contribution of higher sub-bands are ignored. The validity of this assumption is evident from table II.
Combining (\ref{Enk}), (\ref{Q_{sc}}) and (\ref{Q}) with (\ref{E_{fn}}), a relation between $Q_{sc}$, $ \psi_0 $ and $V_{ch}$ is obtained:
\begin{figure}[t]
                \centering
                \includegraphics[width=1\columnwidth]{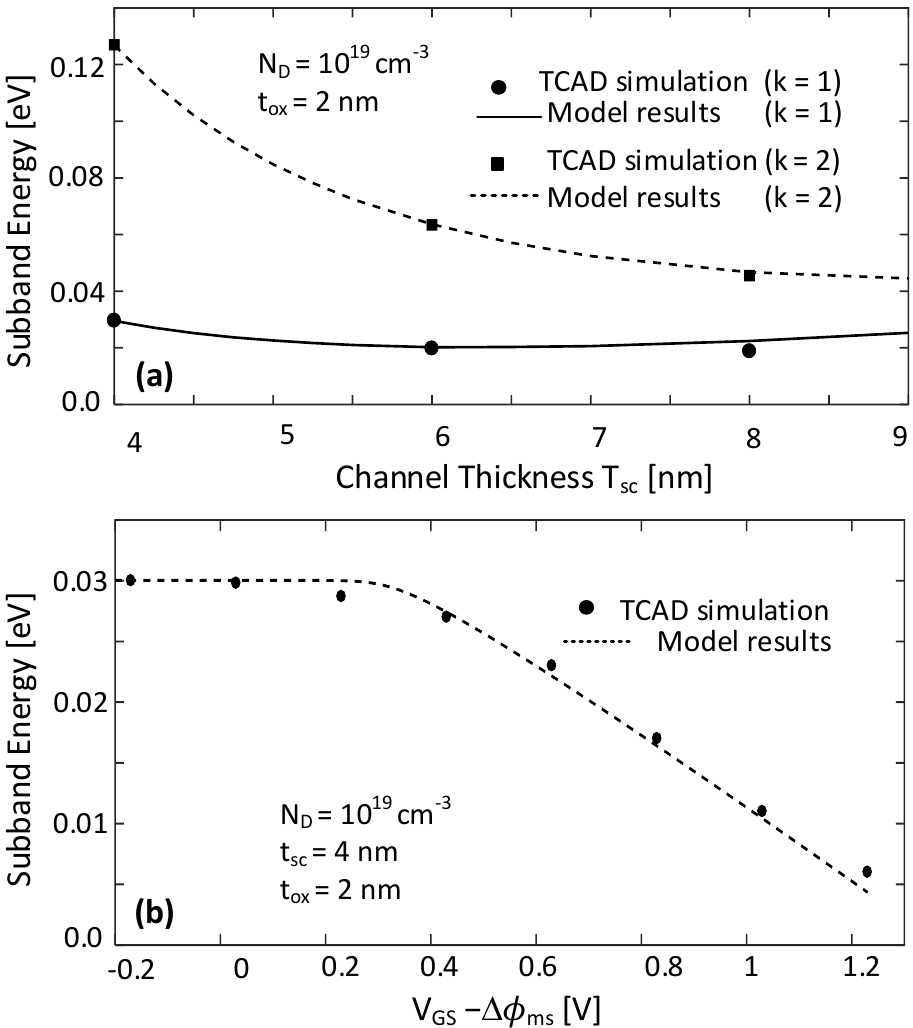}
                \caption{ {\color{black} (a) Lowest Subband energy level as a function of the channel thickness $ T_{sc} $ in the subthreshold region calculated from model and extracted from TCAD simulation  for two degenerated valley (k=1,2), (b) Lowest subband energy as a function of effective gate voltage calculated from the model and extracted from TCAD simulation for one degenerate valley. $ N_D$ = \SI{E19}{cm^{-3}}, $ t_{sc}$ = \SI{4}{nm}, $ t_{ox}$ = \SI{2}{nm}, $ W $ = \SI{1}{\micro m}, $ L_G$ = \SI{1}{\micro m}, $ \mu_n$ = \SI{1100}{cm^2/V\,s}, and $ V_{ch}$ = \SI{0}{V} }.         
                }
                \label{Fig4}
\end{figure}
\begin{equation} \label{Q_sc_sie_0_2}
\begin{split}
&Q_{sc}=qN_DT_{sc}-q\sum_{n=1}^{N_t}\sum_{k=1}^{2}g_{k}N_{k}\times \ln\Bigg\{1+\dfrac{n_i}{N_c}\times\exp \!\\
&\left(\!-\dfrac{(n\pi\hbar)^2}{2qm^*_{c,k}T^2_{sc}U_T}\!-\!\dfrac{Q_{sc}}{24 C_{sc}U_T}\!\left[1\!-\!\dfrac{6}{(n\pi)^2}\right]\!+\!\dfrac{\psi_{0}\!-\!V_{ch}}{U_T}\!\right)\Bigg\}.
\end{split}
\end{equation}\\
Here, it is interesting to highlight that once the center potential $ \psi_0 $ is known, solving relation (\ref{Q_sc_sie_0_2}) gives the total charge density in the channel, and therefore the mobile charge density. As in \cite{5872019, Book}, the center potential still plays a key role in the model.
 
The next step is to introduce the Poisson equation in the model. To end with analytical expressions, we assume that the wave functions satisfy the Schr\"{o}dinger equation for the infinite square quantum well, i.e. neglecting the band bending between the gates. This is a zero order perturbation in the wavefunction (still, the energy was corrected to the first order). As it will be evidenced later, this introduces negligible errors.
In addition, for device thicknesses below \SI {8}{nm} we will consider only the two lowest sub-bands. To check the validity of this assumption, we calculated the percentage of the charge density including the first 10 sub-bands at the flat band condition ($ Q_{SC}=0 $) and found that  the relative error in the carrier density is less than 6\% for the \SI{8}{nm} channel thickness when using relations (\ref{Enk}) and (\ref{Q}) (see table II). The mobile charge distribution obtained by weighting the first and second subband carrier densities $Q_{b1}$ and $Q_{b2}$ by the probability given by the wavefunctions  $w_{1}(x)$ and $w_{2}(x)$ \cite{777154} writes
\begin{equation} \label{ro} \rho (x) =Q_{b1}\left|w_1(x)\right|^2 + Q_{b2}\left|w(x)\right|^2, \end{equation} where $w_n (x)$ is the wavefunction of $ n^{th} $ energy state given by \cite {harrison2016quantum}
\begin{equation} \label{w}
w_{n}(x)=\sqrt{\dfrac{2}{T_{sc}}}\sin\left[\dfrac{n\pi}{T_{sc}}\left(x+\dfrac{T_{sc}}{2}\right)\right].
\end{equation}
Recalling the Poisson equation
\begin{equation} \label{P}
\dfrac{\partial^2}{\partial x^2}\psi(x)=\dfrac{1}{\varepsilon_{si}}\left(\rho- qN_{D}\right),
\end{equation}
and integrating twice using $w_{1,2}(x)$ from (\ref{w}), then  $Q_{b1,2}$ from (\ref{Q}) into (\ref{ro}), leads to an explicit relationship of the potential profile across the gates:
\begin{equation} \label{Sie}
\begin{split}
&\psi(x)=\psi_{0}-\dfrac{1}{2T_{sc}\varepsilon_{si}}\left(\sum_{n=1}^{2}Q_{bn}+q N_D T_{sc}\right)x^2\\ &+\dfrac{Q_{b1}T_{sc}}{4\pi^2\varepsilon_{si}}\left[\cos\left(\dfrac{2\pi x}{T_{sc}}\right)\!-\!1\right]- \dfrac{Q_{b2}T_{sc}}{16\pi^2\varepsilon_{si}}\left[\cos\left(\dfrac{4\pi x}{T_{sc}}\right)\!-\!1\right].
\end{split}
\end{equation}\\
It should be remarked that $ Q_{b1} $, $ Q_{b2} $  are internally linked to ($ \psi_{0} $)  according to (\ref{Q}), meaning that the problem now consists in finding the center potential ($ \psi_{0} $) satisfying (\ref{Q}) and (\ref{Sie}).\\ Importantly, relation (\ref{P}) is not self-consistent with the potential $ \psi(x) $, and in this sense relation (\ref{Sie}) is approximate. \\ As a consequence, the contribution of the depletion charge to the electrostatic potential is discarded in this model. This means that the quadratic dependence of voltages with charges as stated in \cite{5872019, Book} is not included, a simplification which is valid for the very thin channels at aim in this work.
\section{Charge-based approach}
The boundary conditions arising from the continuity of the displacement vector at the interface must
satisfy:
\begin{equation} \label{BC}
V_{GS}-\Delta\phi_{ms}-\psi_{s}=-\dfrac{Q_{sc}}{2C_{ox}},
\end{equation}
where $ \psi_s $ is the potential at the surface of the channel and $ C_{ox}=\varepsilon_{ox}/t_{ox} $  is oxide capacitance per unit area. Solving relation (\ref{Sie}) at the Si-SiO$ _{2} $ interface $ x=\pm T_{sc}/2 $, the surface potential is linked to the center potential through
\begin{equation} \label{SP-CP}
        \psi_{s}-\psi_{0}=-\dfrac{Q_{sc}}{8C_{sc}}-\dfrac{Q_{b1}}{2\pi^2C_{sc}},
\end{equation} 
where $ C_{sc}  =\varepsilon_{si}/T_{sc} $ and $ Q_{b1} $ corresponds to the first sub-band contribution to the mobile charge density, obtained by    
\begin{equation} \label{Q_b1}
Q_{b1}=\beta (Q_{sc}-Q_{fix}),
\end{equation}
where $ \beta $ is the ratio of the first subband mobile charge density to the total mobile charge density, obtained from {\color{black}Table \ref{Table2}}.\\ Finally, merging (\ref{BC}), (\ref{SP-CP}) and (\ref{Q_b1}) links the total charge density to the center potential for a given value of $ V_{GS} $:
\begin{figure}[!t]
                \vspace{0.08cm}
                \centering
                \includegraphics[width=1\columnwidth]{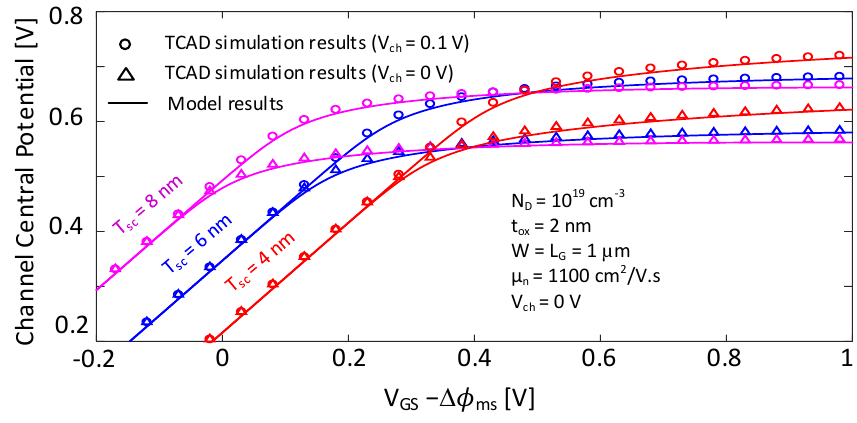}
                \caption{Center potential with respect to the effective gate voltage for different values of $ T_{sc} $ and $ V_{ch} $ = \SI{0}{V} and \SI{0.1}{V}. The proposed model: solid lines and TCAD simulation results, triangle symbols: $ V_{ch} $ = \SI{0}{V} and circle symbols: $V_{ch} $ = \SI{0.1}{V} ($ N_D $ = \SI{E19}{cm^{-3}}, $ t_{ox}$ = \SI{2}{nm}, $ W $ = \SI{1}{\micro m}, $ L_G$ = \SI{1}{\micro m}, $ \mu_n$ = \SI{1100}{cm^2/V\,s}).}
                \label{Fig5}
                \vspace{-0.2cm}
\end{figure}
\begin{equation} \label{Q_sc_sie_0}
V_{GS}\!-\!\Delta \phi_{_{ms}}\!-\!\psi_{0}=-\dfrac{Q_{sc}}{8C_{sc}}-\dfrac{\beta Q_{sc}}{2\pi^2C_{sc}}-\dfrac{Q_{sc}}{2C_{ox}}+\dfrac{\beta Q_{fix}}{2\pi^2C_{sc}}.
\end{equation}\\
\begin{figure}[b]
        \centering
        \includegraphics[width=0.9\columnwidth]{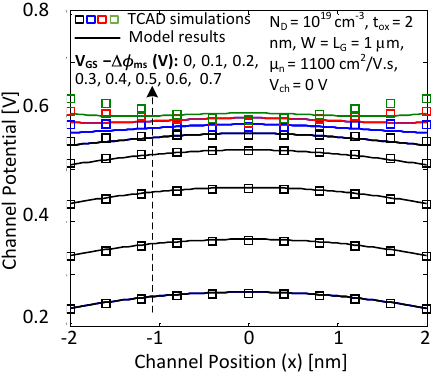}
        \caption{Potential profile across the channel for $ T_{sc} $ = \SI{4}{nm} and different values of the effective gate voltage. Solid lines and symbols are corresponding to the model results and TCAD Simulation results.}
        \label{Fig6}
        \vspace{-0.2cm}
\end{figure}
\begin{figure}[t]
        \centering
        \includegraphics[width=1\columnwidth]{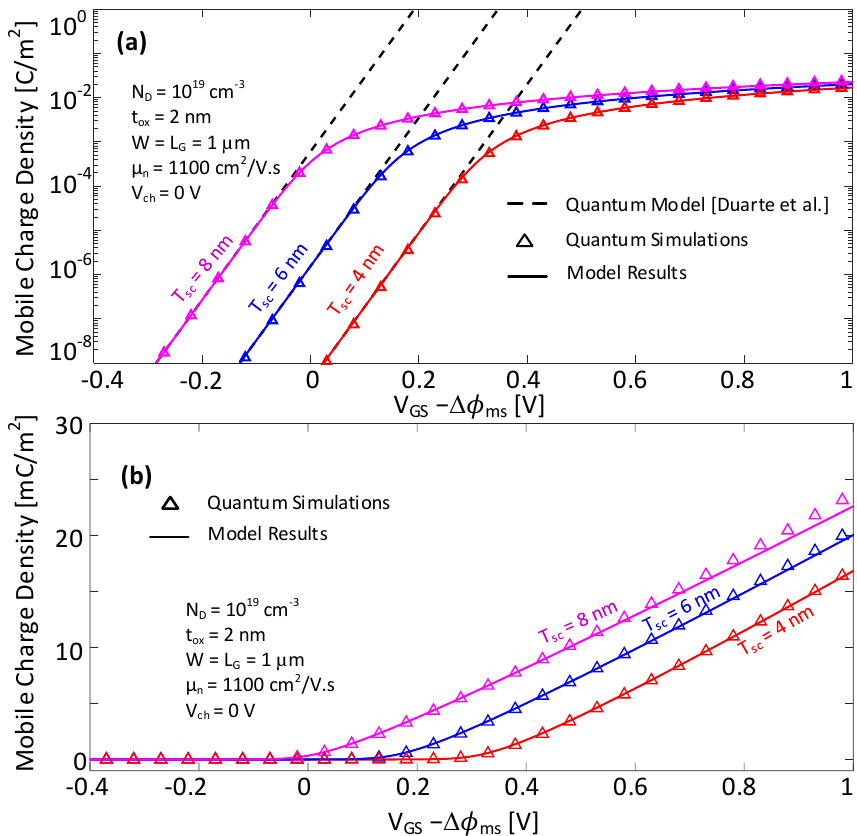}
        \caption{Mobile charge density versus the effective gate voltage for different values of $ T_{sc} $ in logarithmic (a) and linear scales (b). Solid lines, dashed lines and triangle symbols are respectively corresponding to the quantum model, quantum TCAD simulation and the model  proposed by Duarte et al. ( $ N_D$ = \SI{E19}{cm^{-3}}, $ t_{ox}$ = \SI{2}{nm}, $ W $ = \SI{1}{\micro m}, $ L_G$ = \SI{1}{\micro m}, $ \mu_n$ = \SI{1100}{cm^2/V\,s}, and $ V_{ch}$ = \SI{0}{V}).               
        }      
        \label{Fig7}
        \vspace{-0.4cm}
\end{figure}
For fixed $ V_{GS} $ and quasi Fermi potentials ($ V_{ch} $), relations (\ref{Q_sc_sie_0_2}) and (\ref{Q_sc_sie_0}) can be solved to obtain $ \psi_0 $ and $ Q_{sc} $.
The channel center potential is plotted at the source ($ V_{ch}=0V $) and the drain ($ V_{ch}=0.1V $) with respect to the effective gate voltage in Fig. \ref{Fig5}. For \SI{4}{nm} channel thickness, flat-band occurs at $ V_{GS}\approx$ \SI{0.54}{V}.
 
Fig. \ref{Fig6} compares the potential distribution across the channel for different values of $ V_{GS} $. The model is accurate with respect to numerical  simulation up to the flat-band voltage. For higher gate voltages, a small deviation is evidenced, most likely because of the large charge density that would ask for an additional correction of the wavefunctions (unlike for the energy states, wavefunctions are the solutions of the infinite square quantum well). The mobile charge density versus $ V_{GS} $ for different channel thicknesses is depicted on Fig. \ref{Fig7}. The model matches TCAD simulations in all the regions of operation, both in linear and logarithmic scales (note that the derivation in \cite{6151818} is accurate in deep depletion only).
An approximation of the center potential $ \psi_{0} $ can be further obtained from (\ref{Q_sc_sie_0_2}) when using the Boltzmann instead of the Fermi-Dirac statistics
\begin{equation} \label{psi_FB}
\begin{split}
\psi_{0}\!=&\!V_{ch}\!+U_T\log\left(qN_D T_{sc}-Q_{sc}\right)\\
&-\!U_T\log\left[q\sum_{n=1}^{N_t}\!\sum_{k=1}^{2}g_{k}N_{k}\dfrac{n_i}{N_c} \exp\left(-{\dfrac{E_{n,k}}{U_T}}\right)\right].
\end{split}
\end{equation}
It should be noted that at flat band, i.e. $(Q_{sc}=0)$, $ \psi_{0, FB} $ depends on the thickness of the channel as indicated in Fig. \ref{Fig5} (in contrast to the classical limit where $\psi_{0,FB}(classic)=V_{ch}+U_T\log({N_D}/{n_i})$ because quantum confinement still shifts the ground state in the channel and affects the flat band potential.
If only the mobile charge density of the first subband $(n=1)$ is considered, while neglecting contributions from higher sub-bands $(n>1)$, relation (\ref{Q_sc_sie_0_2}) can be simplified:
\begin{equation} \label{17}
\psi_{0}-V_{ch}-\dfrac{Q_{sc}}{24C_{sc}}\left(1-\dfrac{6}{\pi^2}\right)=U_T\ln \left(\dfrac{Q_{fix}-Q_{sc}}{Q_0}\right),
\end{equation}
where $ Q_0 $ is given by
\begin{equation} \label{Q0}
Q_0=\dfrac{qk_BT}{\pi \hbar^2}.\dfrac{n_i}{N_c}\sum_{k}^{2} g_k m^*_{d,k} \exp\left[\!-\dfrac{(\pi\hbar)^2}{2k_BTm^*_{c,1}T^2_{sc}}\right].
\end{equation}\\
Next, replacing $ \psi_{0} $ from (\ref{17}) in (\ref{Q_sc_sie_0}) leads to
\begin{equation} \label{19}
\begin{split}
V_{GS}-\!&\Delta\phi_{ms}\!-\!V_{ch}\!-\!\dfrac{\beta Q_{fix}}{2\pi^2 C_{sc}}\!-\!U_T\ln\left(\!\dfrac{Q_{fix}}{Q_0}\!\right)\!\\
&=\!U_T\!\ln\left(\!1\!-\!\dfrac{Q_{sc}}{Q_{fix}}\!\right)  -\alpha.Q_{sc},
\end{split}
\end{equation}
\begin{equation} \label{alpha}
\alpha=\left[\dfrac{1}{8C_{sc}}\!+\!\dfrac{\beta}{2\pi^2C_{sc}}\!+\!\dfrac{1}{2C_{ox}}\!-\!\dfrac{1}{24C_{sc}}\left(1\!-\!\dfrac{6}{\pi^2}\right)\right].
\end{equation}\\
Fig. \ref{Fig7} confirms that the mobile charge density calculated from (\ref{19}) with (\ref{Q_sc_sie_0_2}) and (\ref{Q_sc_sie_0}) is accurate for channel thicknesses of \SI{4}{nm} and \SI{6}{nm}. Concerning the 8nm-channel thickness, neglecting the second subband underestimates the mobile charge density in accumulation.
Furthermore, for $T_{sc}$ greater than $8$ nm, infinite quantum well with first order perturbation theory overestimates the first subband as illustrated in Fig. \ref {Fig4}
(a). The electrostatic potential becomes more effective on the energy subbands} \cite{6151818} and either quantum harmonic oscillator (QHO) for subthreshold region or triangular QW for accumulation mode provide  better approximations of the Schrödinger equation. 

\section{Drain Current Derivation}
Relying on the drift-diffusion transport model, the total current is given by
\begin{equation} \label{Current}
I_{DS}=\dfrac{W}{L_G}\mu_n Q_{fix}-\dfrac{W}{L_G}\mu_n\int\limits_{S}^{D} Q_{sc}dV_{ch},
\end{equation}
where $ \mu $ is the free carrier mobility and $ W $ is corresponding to the Width of the device. The carrier mobility is assumed constant along the channel for the sake of simplicity.
 
\begin{figure}[t]
         \centering
         \includegraphics[width=1\columnwidth]{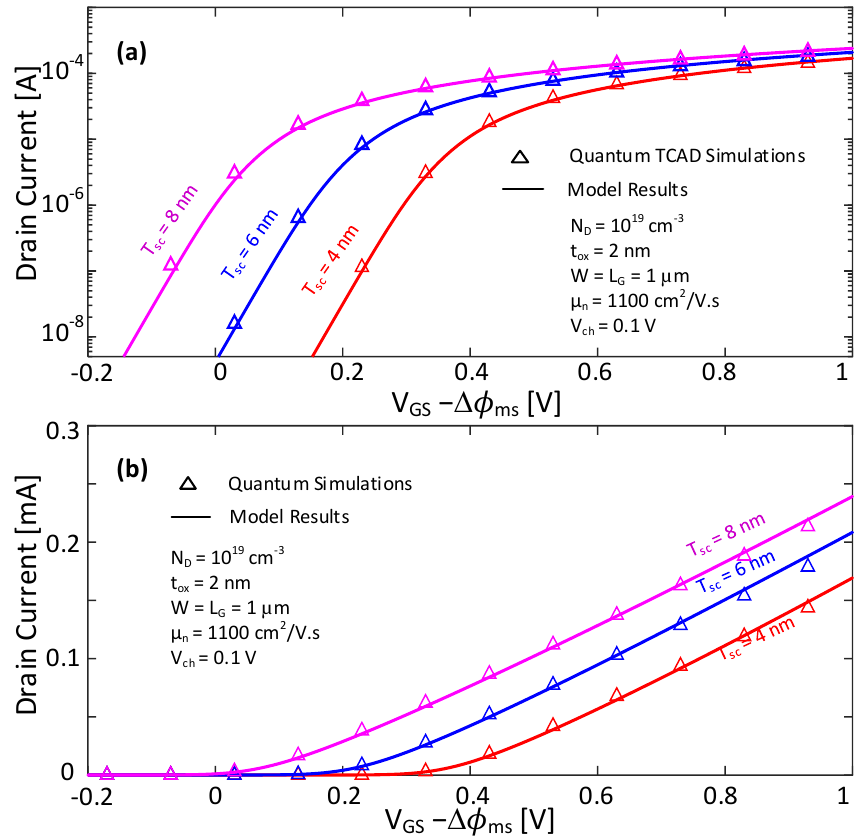}
         \caption{Drain current versus the effective gate voltage for $ V_{DS}= $\SI{0.1}{V} and different channel thicknesses in logarithmic (a) and linear scales (b). Solid lines  and triangle symbols are respectively corresponding to the proposed quantum model and  TCAD simulation results ( $ N_D$ = \SI{E19}{cm^{-3}}, $ t_{ox}$ = \SI{2}{nm}, $ W $ = \SI{1}{\micro m}, $ L_G$ = \SI{1}{\micro m}, $ \mu_n$ = \SI{1100}{cm^2/V\,s}).    }
         \label{Fig9}
         \vspace{-0.2cm}
\end{figure}

\begin{figure}[t]
         \centering
         \includegraphics[width=1\columnwidth]{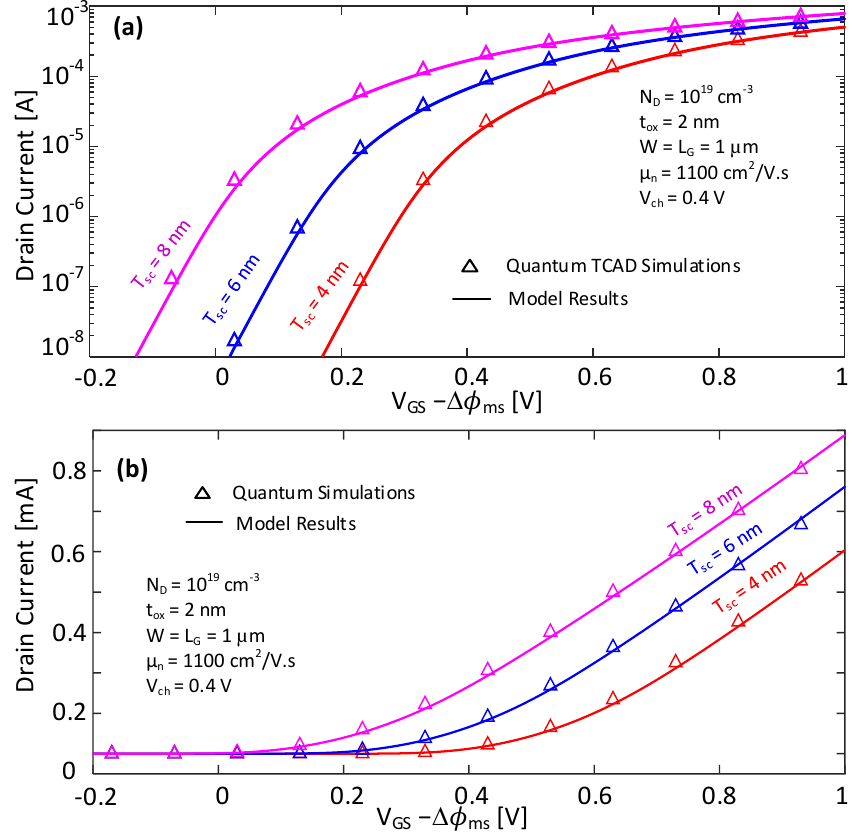}
         \caption{Drain current versus gate voltage for $ V_{DS}= $\SI{0.4}{V} and for Channel thickness of 4, 6 and 8 nm. Solid lines and triangle symbols are respectively corresponding to the proposed model and  TCAD simulation results ( $ N_D$ = \SI{E19}{cm^{-3}}, $ t_{ox}$ = \SI{2}{nm}, $ W $ = \SI{1}{\micro m}, $ L_G$ = \SI{1}{\micro m}, $ \mu_n$ = \SI{1100}{cm^2/V\,s}).}
         \label{Fig10}
         \vspace{-0.2cm}
\end{figure}

Differentiating (\ref{19}) gives $ dV_{ch} $ in terms of $ Q_{sc} $ and $ dQ_{sc} $;
\begin{equation} \label{dV-dQ}
dV_{ch}=\alpha dQ_{sc}+U_TdQ_{sc}\left(1-\dfrac{Q_{sc}}{Q_{fix}}\right)^{-1},
\end{equation} 
Multiplying both side of (\ref{dV-dQ}) with $Q_{sc} $ and integrating once leads to
\begin{equation} \label{22}
\begin{split}
\!\int\limits_{S}^{D}Q_{sc}dV_{ch}\!=\!\left.\dfrac{\alpha}{2}Q^2_{sc}\right|^D_S\!-\!\left.U_TQ_{sc}\right|^D_S\!-\!\left.U_TQ_{fix}\ln\left(1\!-\!\dfrac{Q_{sc}}{Q_{fix}}\right)\right|^D_S
\end{split}
\end{equation}
Finally, introducing (\ref{22}) in (\ref{Current}), the total current is obtained from the semiconductor charge densities evaluated at source and drain.
The current versus gate voltage is plotted in Fig. \ref{Fig9} and Fig. \ref{Fig10} for $ V_{DS}=0.1 V $ and $ V_{DS}=0.4 V $ respectively. The length and the width of the device are $ 1 \mu m $ to avoid the short and narrow channel effects. The electron mobility was set to a constant value of  \SI{0.11}{m^2/Vs}. Bohm quantum potential (BQP) model calibrated with Poisson-Schr\"{o}dinger model were used to predict electron transport in TCAD simulations. Using the same parameters as for TCAD simulations (i.e. no empirical parameters), the model predicts the current characteristics in all regions of operation with an accuracy of 93\%.
In the classical limit, when the thickness of the channel increases so that the energy between subbands is comparable with $ k_B T $, the series in (\ref{psi_FB}) can be replaced by integral using the identity
\begin{equation} \label{23}
\sum_{n}^{}\exp\left[-\left(\sigma n\right)^2\right]=\int\limits_{0}^{\inf}\exp\left[-\left(\sigma.x\right)^2\right]dx=\dfrac{\sqrt\pi}{2\sigma},
\end{equation}
where $ \sigma =\pi\hbar/\left(T_{sc}\sqrt{2m^*_{c,k}k_BT}\right) $. Then, at the flat band condition ($ Q_{SC}=0 $), the dependence on $ T_{sc} $ cancels in (\ref{psi_FB}) and the classical definition of $ \psi_{0, FB} $  is recovered. 
\section{Conclusion}\label{sec:conclusion}
We derived and discussed a charge based model for ultra-thin JL DG FET which takes into account charge quantization in all the regions of operation. The analytical solution introduces a zero order approximation in wave-functions and a first order correction for the confined energies. The model was validated with TCAD simulation for devices with channel thicknesses of 8 nm down to 4nm and shows high accuracy. This approach could be used as a core model for JLFETs implemented in Ultra Thin Body SOI technology.    
\appendices
\bibliographystyle{IEEEtran}
\bibliography{IEEEabrv,biblio}
\vspace{-4cm}
\begin{IEEEbiography}[{\includegraphics[width=1in,height=1.25in,clip,keepaspectratio]{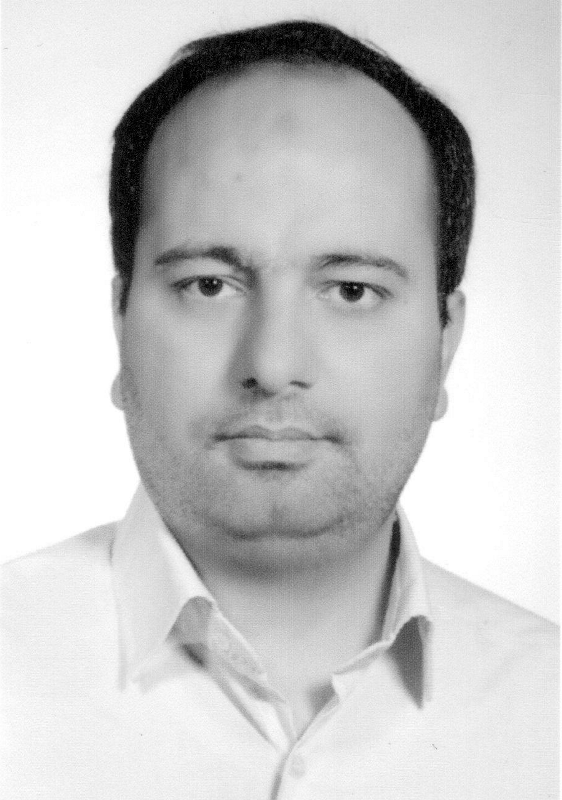}}]{Majid Shalchian} received the Ph.D. degree in electrical engineering (microelectronics) from Sharif University of Technology, Tehran, IRAN, in 2005.He  joined Amirkabir University of Technology (Tehran Polytechnique), as assistant professor in electrical engineering department. His current research interests include semiconductor device modelling and parameter extraction  as well as VLSI circuit design. He is the director of VLSI Laboratory and advanced automotive electronics Laboratory in Amirkabir University of Technology.  
\end{IEEEbiography}

\vspace{-3cm}

\begin{IEEEbiography}[{\includegraphics[width=1in,height=1.25in,clip,keepaspectratio]{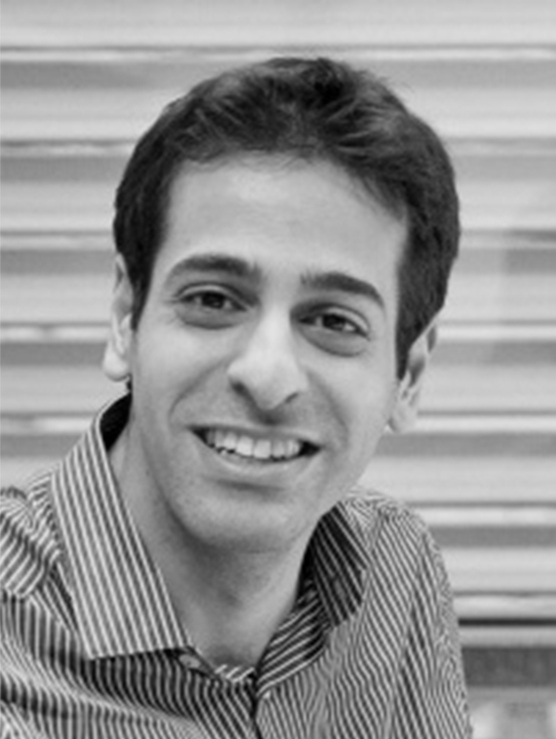}}]{Farzan Jazaeri} received the Ph.D. degree in microelec tronics and microsystems from the Ecole Polytechnique Fédérale de Lausanne (EPFL), Lausanne, Switzerland, in 2015. He joined the Integrated Circuits Laboratory, EPFL, as a Research Scientist and a Project Leader. His current research interests include solid state physics and advanced semiconductor devices for operation within extreme harsh environments, i.e., high energy particle background and cryogenic temperatures for space applications and quantum computations.
\end{IEEEbiography}

\vspace{-3cm}

\begin{IEEEbiography}[{\includegraphics[width=1in,height=1.25in,clip,keepaspectratio]{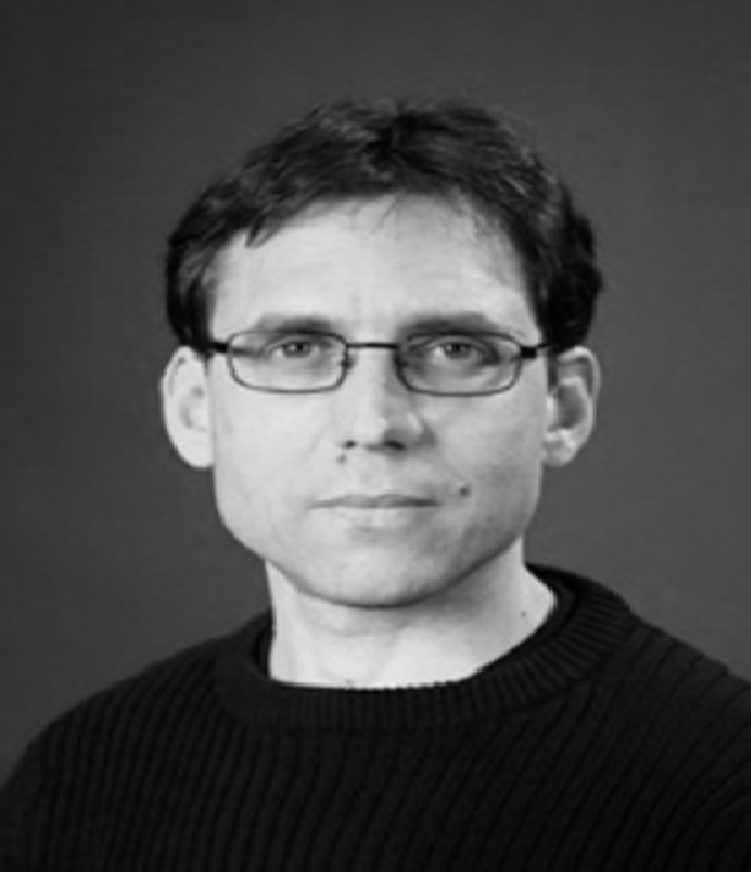}}]{Jean-Michel Sallese} received the Ph.D. degree in physics from the University of Nice Sophia Antipolis, Nice, France. He joined the Ecole Polytechnique Fédérale de Lausanne, Lausanne, Switzerland, and was appointed as the Maitre d’Enseignement et de Recherche. His group (Electron Device Modeling and Technology Laboratory ) has dedicated attention to the solid state physics and modeling of field-effect transistors, radiation damages in integrated circuits collaboration with CERN, and heterostructure HEMT devices.
	
\end{IEEEbiography}

\end{document}